\documentclass[conference]{IEEEtran}

 \usepackage{multirow}
 \usepackage[table,xcdraw]{xcolor}

\ifCLASSINFOpdf
   \usepackage[pdftex]{graphicx}
\else
\fi
\usepackage{amsmath}
\begin{document}
%
\title{The Impact of Blocking Cars on Pathloss Within \\a Platoon: Measurements for 26 GHz Band}


\author{
\IEEEauthorblockN{Pawe\l~Kryszkiewicz}
\hspace{-1.5cm}\IEEEauthorblockA{\textit{\small Poznan University of Technology}\\
\small Poznan, Poland \\
\small pawel.kryszkiewicz@put.poznan.pl}
\and
\IEEEauthorblockN{Adrian~Kliks}
\hspace{-1.5cm}\IEEEauthorblockA{\textit{\small Poznan University of Technology}\\
\small Poznan, Poland \\
\small adrian.kliks@put.poznan.pl}
\and
\IEEEauthorblockN{Pawe\l~Sroka}
\IEEEauthorblockA{\textit{\small Poznan University of Technology}\\
\small Poznan, Poland \\
\small pawel.sroka@put.poznan.pl}
\and
\IEEEauthorblockN{Micha{l}~Sybis}
\hspace{-1.5cm}\IEEEauthorblockA{\textit{\small Poznan University of Technology}\\
\small Poznan, Poland \\
\small michal.sybis@put.poznan.pl}
}


%


\maketitle

\begin{abstract}
Platooning is considered to be one of the possible prospective implementations of the autonomous driving concept, where the train-of-cars moves together following the platoon leader's commands. However, the practical realization of this scheme assumes the use of reliable communications between platoon members. In this paper, the results of the measurement experiment have been presented showing the impact of the blocking cars on the signal attenuation. The tests have been carried out for the high-frequency band, i.e. for 26.555 GHz. It has been observed that on one hand side, the attenuation can reach even tens of dB for 2 or 3 blocking cars, but in some locations, the impact of a two-ray propagation mitigates the presence of obstructing vehicles. 
\end{abstract}


%
\IEEEpeerreviewmaketitle

\section{Introduction}
For many years the safety and reliability of moving cars was the driving force of the automotive industry development. Various technical inventions became an inherent part of nowadays cars increasing the level of comfort and safety. One may include for example Anti-lock Braking Systems (ABS), Anti Slip Regulation (ASR), Electronic Stability Program (ESP), but also lane assist systems, pre-collision assist or Adaptive Cruise Control (ACC) system, just to mention a few of them. 
Nowadays, this branch of the economy is entering into its new phase - the time, when cars will be with high probability driven autonomously with the support of artificial intelligence.
One of the simplest realizations of the self driving concept is the so called autonomous platooning, presented in, e.g., \cite{Zheng2016, Ucar2018}. In a nutshell, a group of vehicles that drive together in a coordinated manner, typically led by a~leader, can be referred to as a platoon. Such a car structure can be created in advance (i.e., at the beginning of the travel), or dynamically on a~road. In the second case, cars create "the road-train" on-demand, and at any time the new car can join the platoon, and any platoon member may move out. Numerous researches have proved that the use of a~platoon has various benefits, including an increase in road capacity \cite{LPT+16}, reduction of fuel consumption, and, consequently, lower carbon footprint~\cite{SARTRE}. These goals will be achieved when the platoon members will drive very close to each other, i.e., the inter-car distance within the platoon is minimized. One of the prospective supportive solutions is the so-called Cooperative Adaptive Cruise Control (CACC) \cite{Dey2016}. In this scheme, information gathered based on readings of the on-board sensors is enriched with data exchange between vehicles using wireless communications. Precise, reliable, and fast communications between the cars has to be guaranteed.
Such a communications system can be typically realized based on either dedicated short-range communications (DSRC) or Cellular Vehicular-to-Everything (C-V2X) system. In the former one, the two lower layers are realized based on the IEEE 802.11p standard \cite{IEEE80211}, whereas, in the case of the second one, an extension of the Long Term Evolution (LTE) or New Radio (NR) standards are applied \cite{VUKAD2018}. It has been shown that both these technologies are sensitive to the wireless channel congestion problem, i.e., to the situation when a high number of simultaneously transmitting vehicles is experienced \cite{VUKAD2018}. 
In particular, both steering information multicast by the platoon leader, as well as data exchanged between the neighboring cars has to be received and decoded correctly for keeping platoon crash-free movement. 
Various solutions can be implemented to improve the signal quality, measured typically by  Signal-to-Interference-plus-Noise Ratio (SINR), such as advanced channel coding or the application of sophisticated reception algorithms. One of the promising solutions is to shift the transmission within the platoon to other bands, which tend to be less occupied, such as vacant TV channels or millimeter wave \cite{Sroka2020a}. The usage of higher frequency bands in also considered in the newest IEEE and 3GPP standards, mainly, IEEE 802.11bd and NR V2X, respectively \cite{Naik2019}. In our work, we focus on the latter case, as shifting the traffic from the congested channels to vacant high-frequency bands may be the viable approach for reliability improvement. However, this depends on the interference and propagation properties at a given frequency band. If the traffic is moved to higher frequencies (around 25-40 GHz) a small impact of external interference is expected, as these frequency bands are typically not that heavily occupied as lower (i.e. below 6 GHz) frequency ranges. It is achieved due to the high signal attenuation observed in these frequencies, i.e., when compared to the sub-6 GHz band with the same antenna apertures. 
Thus, the key factors that may have an impact on the signal reception are the distance between the transmitter and the receiver, transmit and receive antenna height as well as the presence of blocking cars on the propagation path. The goal of the conducted experiment was to measure the real impact of these selected factors on the propagation of wireless signals.

In order to model the system correctly and evaluate its performance in a reliable way, the impact of the communication channel has to be precisely considered. Only in such a case, when the applied channel model is solid and well-designed, the conducted computer experiments - especially in the system-level approach \cite{Wang2014} - can provide justified results. Various approaches may be considered for reliable channel modeling, such as those based on ray-tracing/ray-launching schemes, statistical models or utilizing artificial intelligence engine \cite{Hemadeh2018,Wang2018,Zheng2015}. While many models can be found for sub-6 GHz bands, there is a limited number of solutions available for higher-frequency scenarios in the V2V context, with exception for \cite{Huang2020, Li2020}. Thus, in our work, we concentrate on measuring the effective path loss in the platooning scenario while operating on millimeter waves, i.e. we measure the attenuation of the signal due to the presence of cars between the transmitter and receiver. In particular, we have performed static measurements of the path loss as a function of distance, location of the transmit/reception antenna (i.e. the assembly point in the car) as well as a number of cars between the platoon leader and the receiving car. The conducted experiments have been carried out for the centre frequency equal to 26.555 GHz and with the directional antennas. This measurement setup should resemble the setup while using high frequency bands for intra-platoon communications.  

The remaining of the paper is organized as follows. First, we briefly overview existing literature in the context of channel measurements for V2X communications. Next, in Sec.~\ref{sec:expsetup} the experimentation setup is presented and visualized, which is followed by the description of the achieved results in Sec.~\ref{sec:measresults}. The paper is then concluded. 

\section{Channel Modelling for V2X Communications in Platoons }
\label{sec:chanmodoverview}
Various channel models for V2X communications have been considered in the literature. In particular, the propagation within the 5.9 GHz band has been of interest as a consequence of the allocation of this frequency band to IEEE 802.11p-based systems and C-V2X. For example, in \cite{Cheng_5_9GHz_Channel_model}, the authors have conducted a single carrier measurements in a suburban environment in Pittsburgh, PA. They have modeled the pathloss using the so-called single slope and double slope models, but with the addition of supplemental components representing shadowing, described as random variables following a normal distribution.
Next, the authors in \cite{abbas2015measurement} performed the wideband channel sounding measurements, focusing at the 5.6 GHz band. The campaigns have been carried out in both  - urban and highway scenarios - in Sweden.
One of the widely known and used channel models for the 5.9 GHz band is the Geometry-based Efficient propagation model for V2V communication, typically abbreviated as GEMV$^2$ \cite{Boban_TVT_2014}. In this approach, the links are classified as line-of-sight - LOS (Line of Sight), NLOSv (non-LOS because of vehicles), and NLOSb (non-LOS because of buildings). The first one, the LOS link, implements a two-ray ground reflection model, whereas NLOSv takes into account obstacles, and NLOSb applies a log-distance pathloss model with a distance exponent of 2.9. 

When concentrating on the 28 GHz band, the authors of \cite{Solomitckii2020} performed detailed measurements in an urban street canyon scenario and focused on evaluating the prospective impact of the car blockages. In particular, the presence of car(s) between the transmitter and receiver has been evaluated. It has been shown that the attenuation through the clear windows may reach 2 dB, whereas the application of sun-protective film increases it to 15 dB. Moreover, diffraction over the car degrades the received power even by 24 dB. These results have been compared with the ray-based simulation results. 

Next, in \cite{Boban2019}, the authors performed measurements at four frequency bands, mainly 6.75, 30, 60, and 73 GHz. These have been carried out in urban and highway scenarios, allowing for analysis of the influence of the blocking car on the received signal strength. It was focused on car's size and location relative to transmitter and receiver. It is claimed that the usage of high frequencies for V2X communications will be possible even in the case of blocking vehicles. In this context, it is worth mentioning other works that investigate the inter-vehicle communications using mmWaves, mainly around 60 GHz, as e.g. \cite{Yamamoto_2008_V2V_60GHz}. The authors have applied the uniform theory of diffraction to propose propagation models in non-line of sight case with several intermediate vehicles.  

It is also worth mentioning the selected approaches of channel modeling for V2X communications, as performing detailed measurement campaigns is costly and time-consuming. First, the high-frequency channel realizations are obtained through the  application of the ray-tracing or ray-launching models, where geometrical optics or the so-called uniform theory of diffraction (UTD) is applied. As being highly accurate, they are also characterized by their high complexity. On the other hand, stochastic channel models can be applied, but they may be not well-tuned to the real scenarios. The geometry-based stochastic channel models (GSCMs) have been proposed as being able to guarantee a certain level of accuracy but at the expense of accurate channel parametrization (achieved typically by measurements or detailed ray-based modeling). A synergy of the above-mentioned schemes is presented in \cite{Sadovaya2020}.
A survey of various approaches for pathloss modelling is presented in, e.g., \cite{Va2016,He2020}. Furthermore, regarding LTE and NR applications, the study on evaluation methodology of new vehicle-to-everything has been released \cite{3gpp37885}.

The contribution of our work with respect of the prior papers can be summarized as follows - we concentrate on measurements of the high-frequency channel (around 26 GHz) applicable to communications within the platoon, i.e., when multiple cars are traveling jointly forming a convoy of short-distanced vehicles. We focus on the impact of the consecutive cars following the platoon leader on the observed signal power degradation. Moreover, we check also the influence of the antenna montage point on the observed signal power.

\section{Experimentation Setup}
\label{sec:expsetup}

\begin{figure}[!htb]
\centering
\includegraphics[width=2.5in]{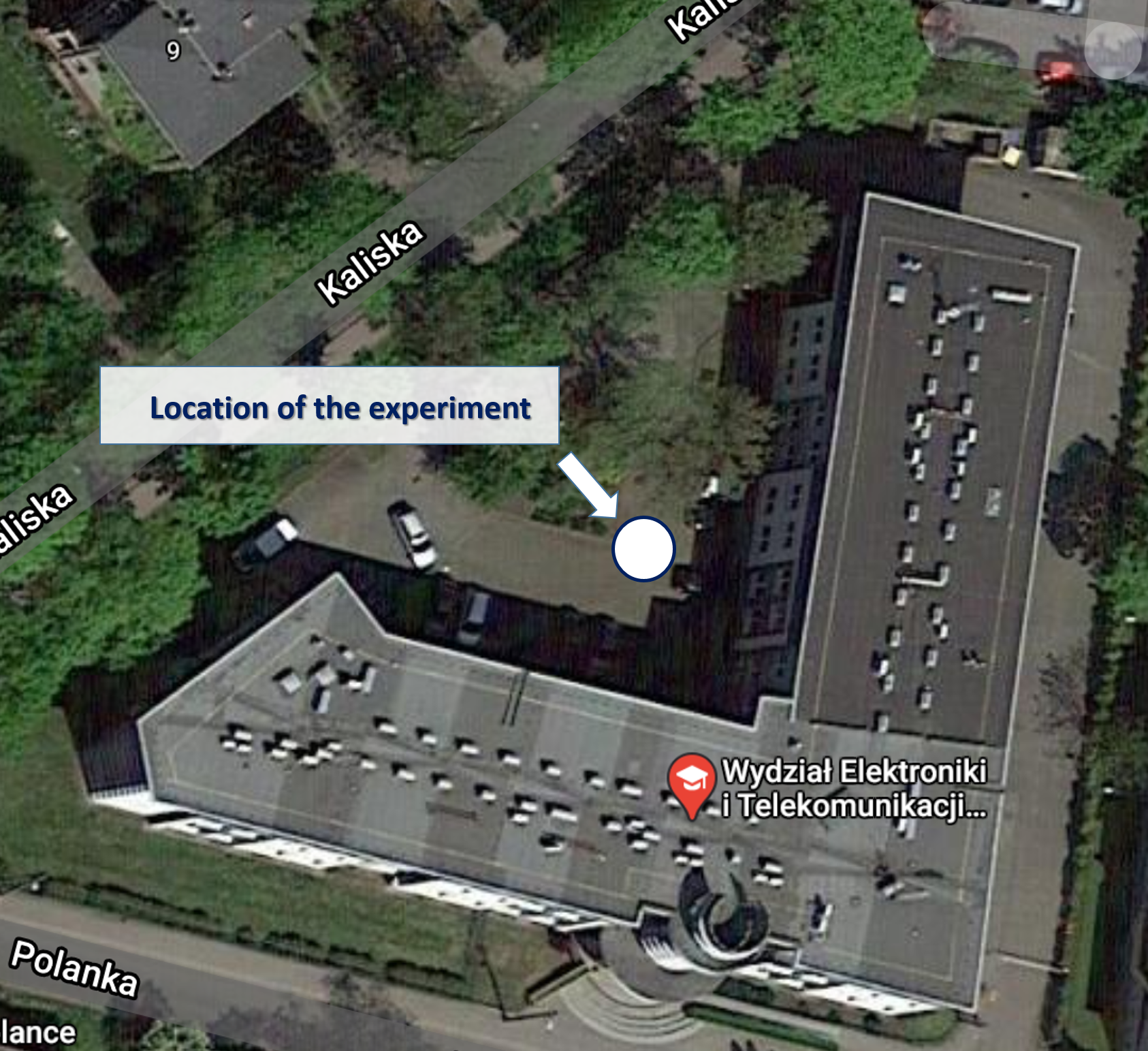}
\caption{Location of the experiment.}
\label{fig_location}
\end{figure}

The measurements have been realized on the 13th of January 2021 at the parking place close to the buildings of the Faculty of Computing and Telecommunications at Poznan University of Technology in Poznan, Poland (coordinates 52°24'01.4"N 16°57'20.3"E), as illustrated in Fig.~\ref{fig_location}. One may observe that around the measurement location, there are just some trees besides the faculty building. 

\begin{figure}[!htb]
\centering
\includegraphics[width=3.4in]{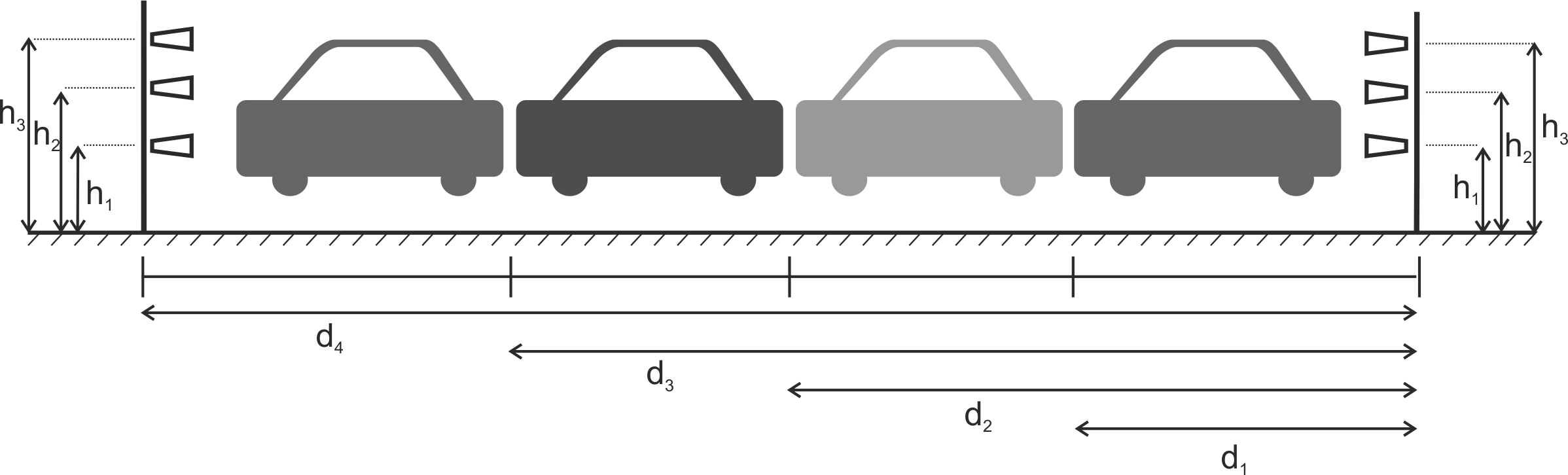}
\caption{Experimental scenarios}
\label{fig_scenarios}
\end{figure}

\begin{figure}[!htb]
\centering
\includegraphics[width=3in]{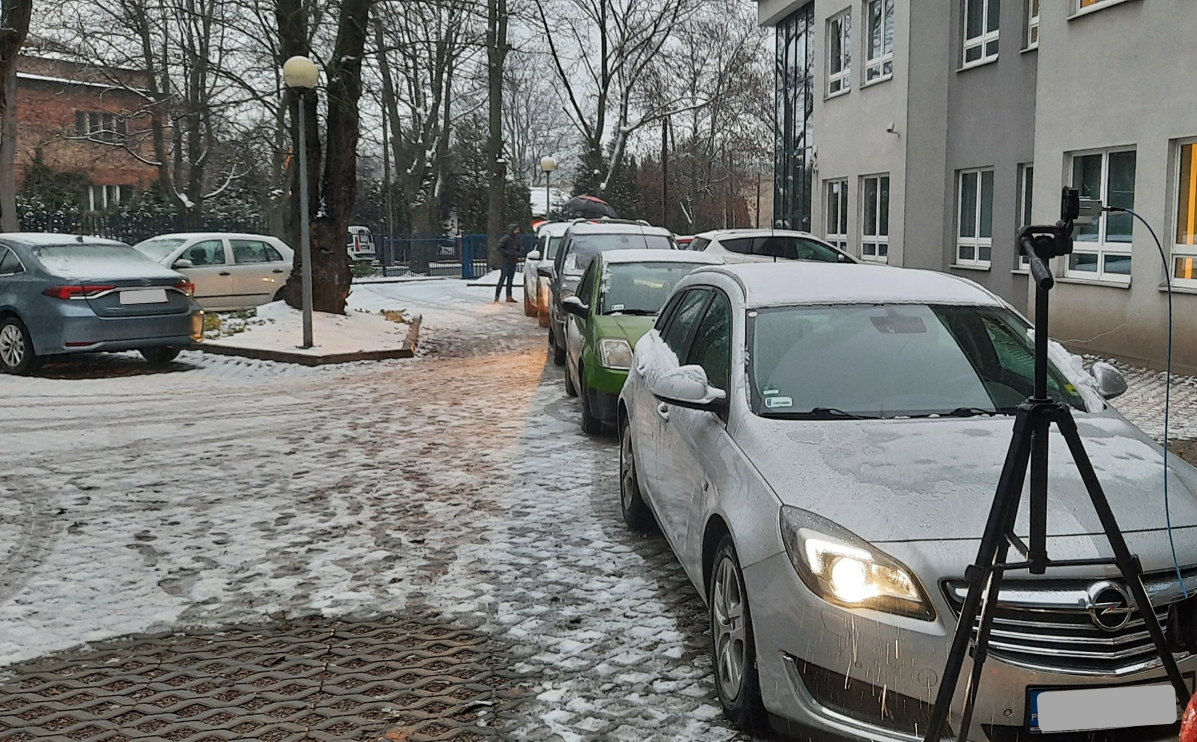}
\caption{Cars deployment during the measurements.}
\label{fig_cars}
\end{figure}

In our experiment, we concentrated strictly on the platooning scenario, when the cars are closely located one by one, and the inter-car distance is very small (down to 2 meters). Our goal was to measure the impact of the wireless channel (its attenuation) on the received signal power by the cars within the platoon. Thus, we wanted to check the prospective signal degradation between the platoon leader (typically the first car in the convoy) and the followers. In particular, we measured how the received signal is attenuated when there is zero, one, two, three, or four cars driving between the transmitter and the receiver (so the platoon size corresponds to the range between two to six). In our scenarios, it corresponds to the following distances between the transmitter and the receiver: 7 m, 12.3 m, 18.2 m, and 24.6 m. Moreover, we have tested three different assembly heights of the antenna, at 0.55 m (i.e., corresponding to the height of car bumper), 1.17 m (i.e., window level), and 1.72 m (car's rooftop). The experimentation setup has been schematically illustrated in Fig.~\ref{fig_scenarios}. Five cars have been selected for this experiment, i.e. Toyota Auris (acting as the transmitter installation point), Opel Insignia, Citroen C2, Peugeot 807, and Hyundai Tucson (these four cars were acting as platoon members, thus were inducing signal attenuation). In Fig.~\ref{fig_cars}, the photography showing the measured platoon has been included. 

During the experiment, the transmitter has been located in the back of Toyota. A single-carrier signal has been generated using the Anritsu MG3694A signal generator. The centre frequency has been fixed to 26.555 GHz, and the transmit power was set to 8 dBm. At the receiver side, frequency spectrum has been observed using Anritsu MS27020T portable spectrum analyzer - the frequency span was set to 2 MHz, and 8-times averaging has been applied. Thus, the results have been stored only when the measurements were stabilized in time.  At both sides of the wireless link, directional antennas Skylink SL-WDPHN-1840-1719-K have been used, with gain of 19.5 dBi and the half-power angle equal to about 15 degrees. These antennas have been mounted on dedicated tripods with regulated height and connected through the 2.92 mm connector and low-loss coaxial cable Jyebao K30K30-53LD-40G150 to the generator and the spectrum analyzer.


\section{Measurements Results}
\label{sec:measresults}
The results have been obtained for five different scenarios, i.e., without any blocking car between the transmitter and receiver (Scenario I, treated hereafter as the reference scenario), and with one to four blocking cars between the transmit antenna and the receive antenna (Scenario II to Scenario V). All the measured values are shown in Tab.~\ref{tab1}. 

\begin{table*}[!t]
\caption{Detailed measurement results.}
\label{tab1}
\begin{tabular}{|p{4cm}|p{2.6cm}|p{2.2cm}|p{2.3cm}|p{2.2cm}|p{2.2cm}|}
\hline
\multicolumn{6}{|c|}{\cellcolor[HTML]{FFE699}\textbf{Measurement setup}} \\
\hline
\multicolumn{1}{|l|}{\textbf{Transmitter setup}} & \multicolumn{5}{|c|}{Single carrier, center frequency  26 555 MHz, P = 8 dBm} \\
\hline
\multicolumn{1}{|l|}{\textbf{Receiver setup}} & \multicolumn{5}{|c|}{Center frequency 26 555 MHz, span 2   MHz, averaging 8x, reference signal level -21 dBm} \\
\hline
\multicolumn{1}{|l|}{\textbf{Cars}} & Toyota Auris & Opel Insignia & Citroen C2 & Peugeot 807 & Hyundai Tucson   \\
\hline
\multicolumn{1}{|l|}{\textbf{Car heights}} &  & 150 & 150 & 175 & 170   \\
\hline
\multicolumn{6}{|c|}{\cellcolor[HTML]{FFE699}\textbf{Scenarios}} \\
\cline{2-6}
  & \multicolumn{1}{|l|}{\cellcolor[HTML]{D9E1F2}d {[}m{]}} & \cellcolor[HTML]{D9E1F2}\textbf{7} & \cellcolor[HTML]{D9E1F2}\textbf{12,3} & \cellcolor[HTML]{D9E1F2}\textbf{18,2} & \cellcolor[HTML]{D9E1F2}\textbf{24,6}  \\
  \cline{2-6}
  & \cellcolor[HTML]{E2EFDA}$P^{I}_1$ for $h$ = 0,55 m & -32,3 dBm& -41 dBm& -41,4 dBm& -56,0 dBm\\
  \cline{2-6}
  &  \cellcolor[HTML]{E2EFDA}$P^{I}_2$ for $h$ = 1,17 m& -36,6 dBm& -36 dBm& -47,6 dBm& -44,2 dBm\\
  \cline{2-6}
  & \cellcolor[HTML]{E2EFDA} $P^{I}_3$ for $h$ = 1,72 m & -37,5 dBm & -46,6 dBm & -50,2 dBm & -52,7 dBm \\
  \cline{2-6}
 \multirow{-5}{*}{Scenario I - lack of blocking cars} & \cellcolor[HTML]{ffc2b3} Average $\mu_1$ (in  linearscale, over \textit{h}) & -34.8 dBm & -39.3 dBm & -44,8 dBm & -48,2 dBm \\
 \hline
  & \multicolumn{1}{|l|}{\cellcolor[HTML]{D9E1F2}d {[}m{]}} & \cellcolor[HTML]{D9E1F2}\textbf{7} & \cellcolor[HTML]{D9E1F2}\textbf{} & \cellcolor[HTML]{D9E1F2}\textbf{} & \cellcolor[HTML]{D9E1F2}\textbf{}  \\
  \cline{2-6}
  & \cellcolor[HTML]{E2EFDA}$P^{II}_1$ for $h$ = 0,55 m & -82,0 dBm &  &  & \\
  \cline{2-6}
  &  \cellcolor[HTML]{E2EFDA}$P^{II}_2$ for $h$ = 1,17 m& -59,9 dBm &  &  & \\
  \cline{2-6}
 &\cellcolor[HTML]{E2EFDA}$P^{II}_3$ for $h$ = 1,72 m & -40,2 dBm &  &  &  \\
  \cline{2-6}
 \multirow{-5}{*}{Scenario II - one blocking car} & \cellcolor[HTML]{ffc2b3} Average $\mu_2$ (in linear scale, over \textit{h}) & -44.9 dBm &  &  & \\
 \hline
   & \multicolumn{1}{|l|}{\cellcolor[HTML]{D9E1F2}d {[}m{]}} & \cellcolor[HTML]{D9E1F2}\textbf{} & \cellcolor[HTML]{D9E1F2}\textbf{12,3} & \cellcolor[HTML]{D9E1F2}\textbf{} & \cellcolor[HTML]{D9E1F2}\textbf{}  \\
  \cline{2-6}
  & \cellcolor[HTML]{E2EFDA}$P^{III}_1$ for $h$ = 0,55 m & &-104 dBm   &  & \\
  \cline{2-6}
  &  \cellcolor[HTML]{E2EFDA} $P^{III}_2$ for $h$ = 1,17 m& &-85 dBm   &  & \\
  \cline{2-6}
  &\cellcolor[HTML]{E2EFDA}$P^{III}_3$ for $h$ = 1,72 m & &-45 dBm   &  &  \\
   \cline{2-6}
 \multirow{-5}{*}{Scenario III - two blocking cars} & \cellcolor[HTML]{ffc2b3} Average $\mu_3$ (in linear scale, over \textit{h}) &  & -49,8 dBm  &  & \\
 \hline
    & \multicolumn{1}{|l|}{\cellcolor[HTML]{D9E1F2}d {[}m{]}} & \cellcolor[HTML]{D9E1F2}\textbf{} & \cellcolor[HTML]{D9E1F2}\textbf{} & \cellcolor[HTML]{D9E1F2}\textbf{18,2} & \cellcolor[HTML]{D9E1F2}\textbf{}  \\
  \cline{2-6}
  & \cellcolor[HTML]{E2EFDA}$P^{IV}_1$ for $h$ = 0,55 m & & &-103 dBm   &  \\
  \cline{2-6}
  &  \cellcolor[HTML]{E2EFDA}$P^{IV}_2$ for $h$ = 1,17 m& & &-94 dBm   &  \\
  \cline{2-6}
  &\cellcolor[HTML]{E2EFDA}$P^{IV}_3$ for $h$ = 1,72 m & & &-71 dBm   &   \\
    \cline{2-6}
 \multirow{-5}{*}{Scenario IV - three blocking cars} & \cellcolor[HTML]{ffc2b3} Average $\mu_4$ (in linear scale, over \textit{h}) &  &  & -75,7 dBm & \\
 \hline
     & \multicolumn{1}{|l|}{\cellcolor[HTML]{D9E1F2}d {[}m{]}} & \cellcolor[HTML]{D9E1F2}\textbf{} & \cellcolor[HTML]{D9E1F2}\textbf{} & \cellcolor[HTML]{D9E1F2}\textbf{} & \cellcolor[HTML]{D9E1F2}\textbf{24,6}  \\
  \cline{2-6}
  & \cellcolor[HTML]{E2EFDA}$P^{V}_1$ for $h$ = 0,55 m & & & &-103 dBm  \\
  \cline{2-6}
  &  \cellcolor[HTML]{E2EFDA}$P^{V}_2$ for $h$ = 1,17 m& & & &-93 dBm \\
  \cline{2-6}
  &\cellcolor[HTML]{E2EFDA}$P^{V}_3$ for $h$ = 1,72 m & & & &-64 dBm  \\
   \cline{2-6}
 \multirow{-5}{*}{Scenario V - four blocking cars} & \cellcolor[HTML]{ffc2b3} Average $\mu_5$ (in linear scale, over h) &  &  &  & -68,8  dBm\\
 \hline
      & \multicolumn{1}{|l|}{\cellcolor[HTML]{D9E1F2}d {[}m{]}} & \cellcolor[HTML]{D9E1F2}\textbf{7} & \cellcolor[HTML]{D9E1F2}\textbf{12,3} & \cellcolor[HTML]{D9E1F2}\textbf{18,2} & \cellcolor[HTML]{D9E1F2}\textbf{24,6}  \\
      \cline{3-6}
      & \multicolumn{1}{|l|}{\cellcolor[HTML]{D9E1F2}} & \cellcolor[HTML]{D9E1F2} \tiny $P^I - P^{II} | \mu_1 - P^{II}$ & \cellcolor[HTML]{D9E1F2} \tiny $P^I-P^{III} | \mu_1 - P^{III}$ & \cellcolor[HTML]{D9E1F2} \tiny $P^I - P^{IV} | \mu_1 - P^{IV}$& \cellcolor[HTML]{D9E1F2} \tiny $P^I - P^V | \mu_1 - P^{V}$ \\
  \cline{2-6}
  & \cellcolor[HTML]{E2EFDA}for $h$ = 0,55 m & 49,7 dB $|$ 47.2 dB &	63,0 dB $|$ 64.7 dB &	61,6 dB $|$ 58.2 dB&	47,0 dB $|$ 54.8 dB \\
  \cline{2-6}
  &  \cellcolor[HTML]{E2EFDA}for $h$ = 1,17 m& 23,3 dB $|$ 25.1 dB&	49,0	dB $|$ 45.7 dB& 46,4 dB $|$ 49.2 dB &	48,8 dB $|$ 44.8 dB\\
  \cline{2-6}
&\cellcolor[HTML]{E2EFDA}for $h$ = 1,72 m & 2,7 dB $|$ 5.4 dB&	-1,6 dB $|$ 5.7 dB&	20,8 dB $|$ 26.2 dB&	11,3 dB $|$ 15.8 dB  \\
    \cline{2-6}
 \multirow{-6}{*}{Calculated attenuation} & \cellcolor[HTML]{ffc2b3} Difference & $\mu_2 - \mu_1 = $ 10.1 dB  & $\mu_3 - \mu_1 =$ 10.5 dB & $\mu_4 - \mu_1$ = 30.9 dB & $\mu_5 - \mu_1$ = 20.6 dB\\
 \hline
\end{tabular}
\end{table*}

\begin{figure}[!htb]
\centering
\includegraphics[width=2.8in]{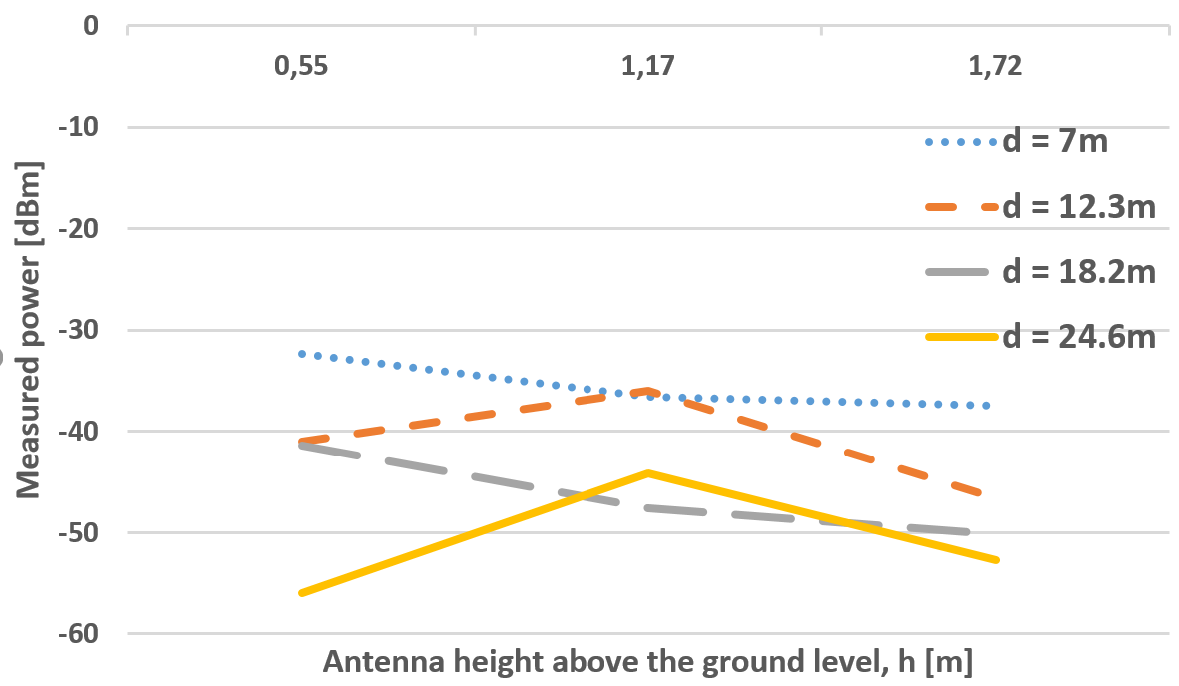}
\caption{Measured power as function of antenna height}
\label{fig_PLheight}
\end{figure}

\begin{figure}[!htb]
\centering
\includegraphics[width=2.8in]{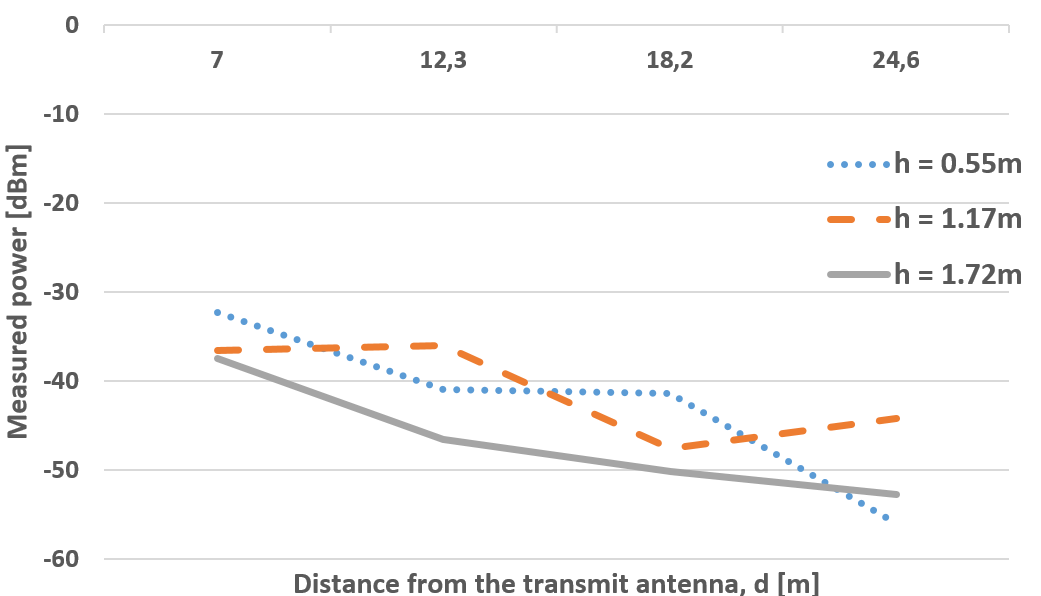}
\caption{Measured power as function of distance from the transmitter}
\label{fig_PLdistance}
\end{figure}

First, in Figs~\ref{fig_PLheight} and \ref{fig_PLdistance}, the measured power has been presented as function of antenna montage height and distance, respectively. These measurements have been achieved for Scenario I. As expected, the pathloss increases as the distance between the transmitter and receiver increases as well. 

\begin{figure}[!htb]
\centering
\includegraphics[width=3in]{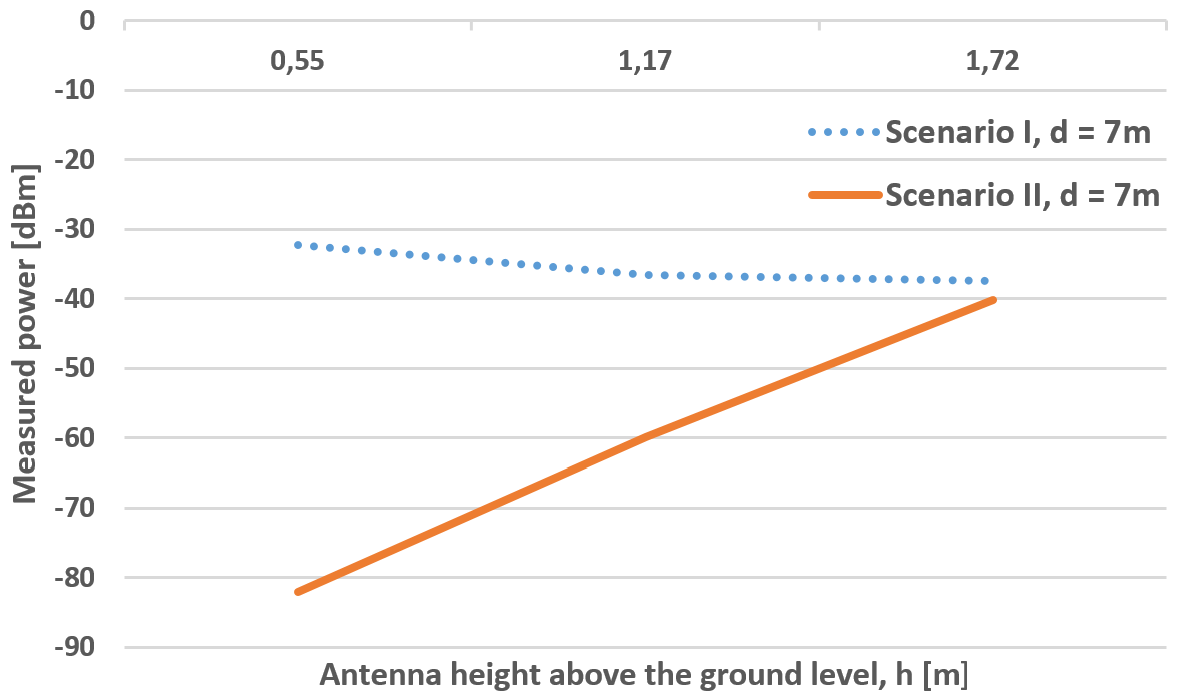}
\caption{Measured power as function of antenna height in Scenarios I and II}
\label{fig_PLone}
\centering
\includegraphics[width=3in]{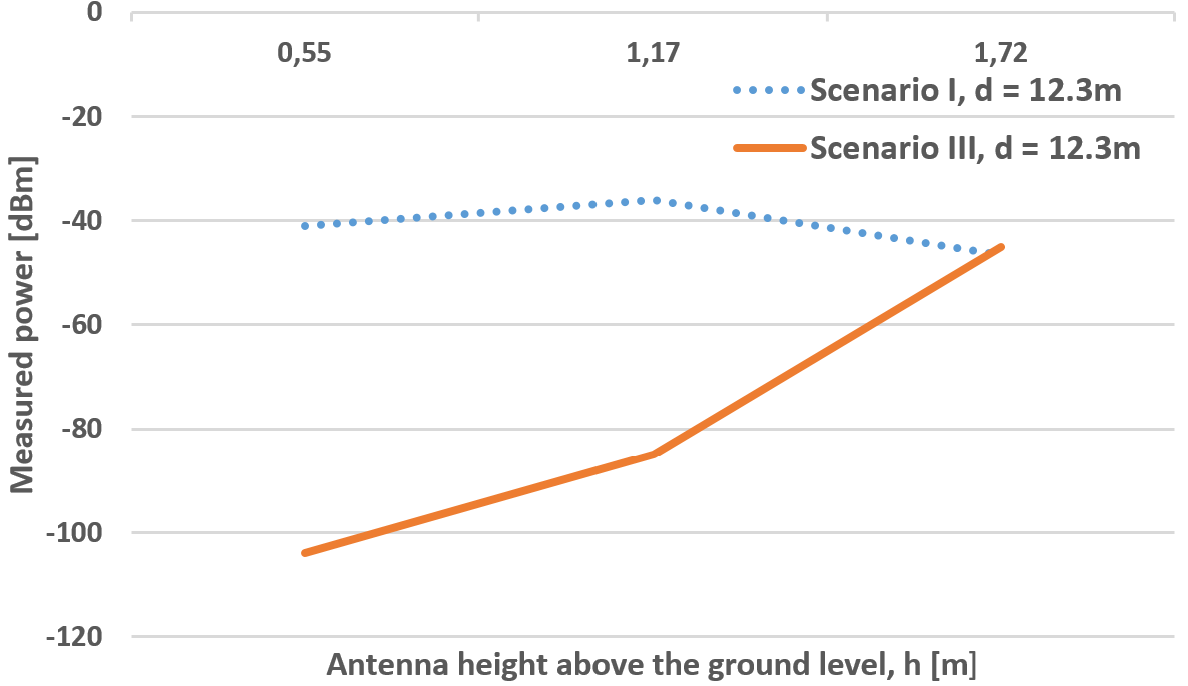}
\caption{Measured power as function of antenna height in Scenarios I and III}
\label{fig_PLtwo}
\end{figure}

\begin{figure}[!htb]
\centering
\includegraphics[width=3in]{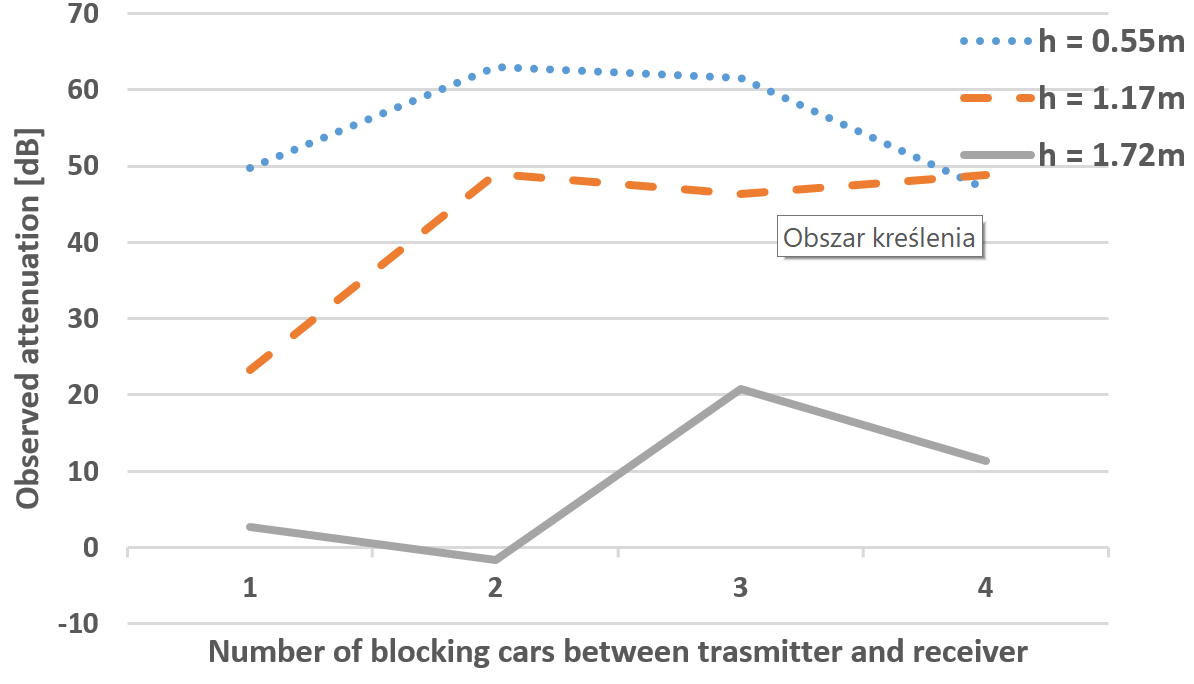}
\caption{Observed attenuation}
\label{fig_attenuation}
\end{figure}

Moreover, in Figs.~\ref{fig_PLone} - \ref{fig_PLfour}, the values of measured power for all five considered scenarios have been compared with the reference case (i.e., when no car was present in the signal path - Scenario I). Finally, attenuation introduced by in-platoon cars has been calculated and presented in Fig.~\ref{fig_attenuation}.

\begin{figure}
\centering
\includegraphics[width=3in]{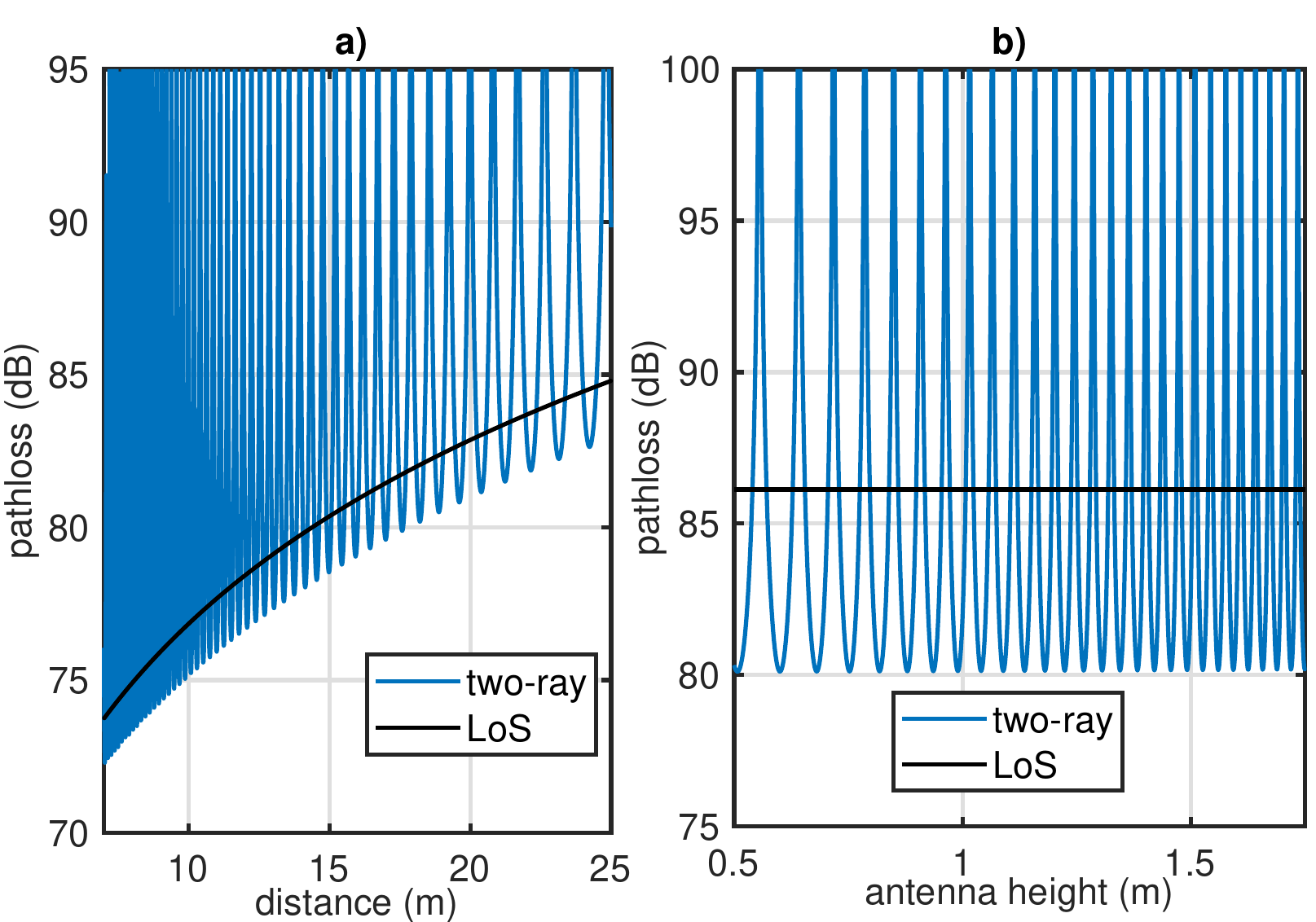}
\caption{Pathloss for two-ray propagation with reflection from ground: a) as a function of distance with antenna height h=1.72m, b) as a function of antenna height for distance d=18.2 m. }
\label{fig_two_ray}
\end{figure}

One can observe that in Scenario I (no blocking cars, i.e., pure free space propagation with a line of sight, possibly with reflection from the ground) the expected increase of signal attenuation as a function of distance can be observed. In particular, the received power for the deployment of antennas at 0.55 m above ground level (agl) decreases from -32.3 dBm at 7 m distance to -56 dBm at 24.6 m. 
However, for some results in Scenario I "anomalies" were observed, e.g., signal power rises for h=1.17 m while increasing the distance from 18.2 m to 24.6 m. This can be justified by the two-ray propagation, i.e., the direct path and the path with reflection from ground adding at the receiving antenna. Examples of theoretical pathloss for such a propagation with ground-reflection coefficient -1 are shown in Fig. \ref{fig_two_ray}. High variations of attenuation called fadings are observed a few times per a single meter in distance and a few times per 10 cm in antenna height. As such, in this scenario the mean received power (in linear scale) over all the measurement heights for a given distance is shown in Tab.~\ref{tab1}. This can be used as an estimate of LoS pathloss for a given distance and used in each next scenario to estimate the attenuation introduced by cars. In the next scenarios the two-ray propagation does not occur, as a result of presence of the cars. As such this averaging is not required. 

After "blocking" the signal by in-platoon car the received signal power typically reduces. While for Scenario II, the observed attenuation due to the presence of one blocking car reaches almost 50 dB for antenna height equal to 0.55 m agl, as for antenna placement at the rooftop there was almost no attenuation increase.
Analogous conclusions can be drawn for Scenario III, IV, and V. An interesting observation is the impact of the antenna height. When the antenna is mounted at the level of the car bumper, the impact of blocking cars located between the transmitter and the receiver is significant. However, when the antenna is at the height of 1.77 m agl, the observed channel attenuation due to the presence of blocking cars is much less severe. On the other hand, some other propagation effects can be observed, as for the distance 12.3 m, the improvement in received signal power is noticed. This is probably the effect of the two-ray propagation, where one path travels below cars and reflects from the ground (i.e. it travels in the tunnel between the ground and the bottom of the car). However, other effects (such as diffraction) may have also impact on the observed results, as discussed e.g. in \cite{Solomitckii2020}.
Finally, interesting observations can be made while analyzing the average attenuation at each distance. Mainly, at each measurement location, the measured logarithmic value has been transformed to a linear scale and averaged over three montage levels. First, one may observe the increase (almost linear in logarithmic scale) of average attenuation as a function of distance - it increases around 5 dB per each 5 - 6 m. Moreover, the difference between the average attenuation in Scenario I and every other scenario has been calculated, and the results are put in Tab.~\ref{tab1} in the last row. One may observe that one or two cars cause the increase of average attenuation for about 10 dB, whereas three or four cars increase average attenuation by 31 and 21 dB, respectively. 
%

\begin{figure}[!htb]
\centering
\includegraphics[width=3in]{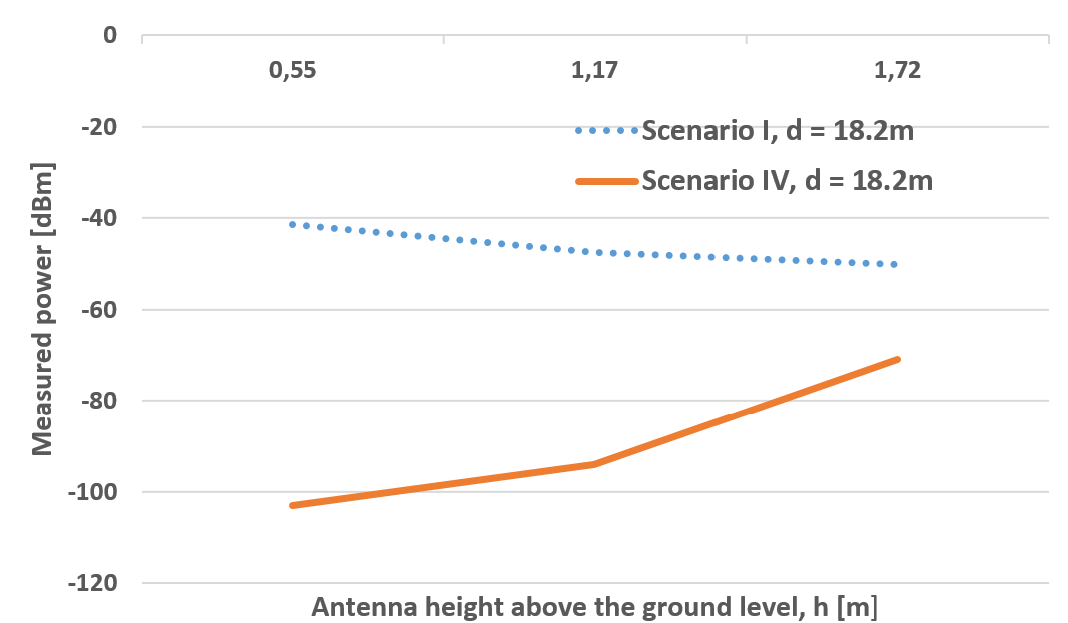}
\caption{Measured power as function of antenna height in Scenarios I and IV}
\label{fig_PLfour}
\end{figure}

\begin{figure}[!htb]
\centering
\includegraphics[width=3in]{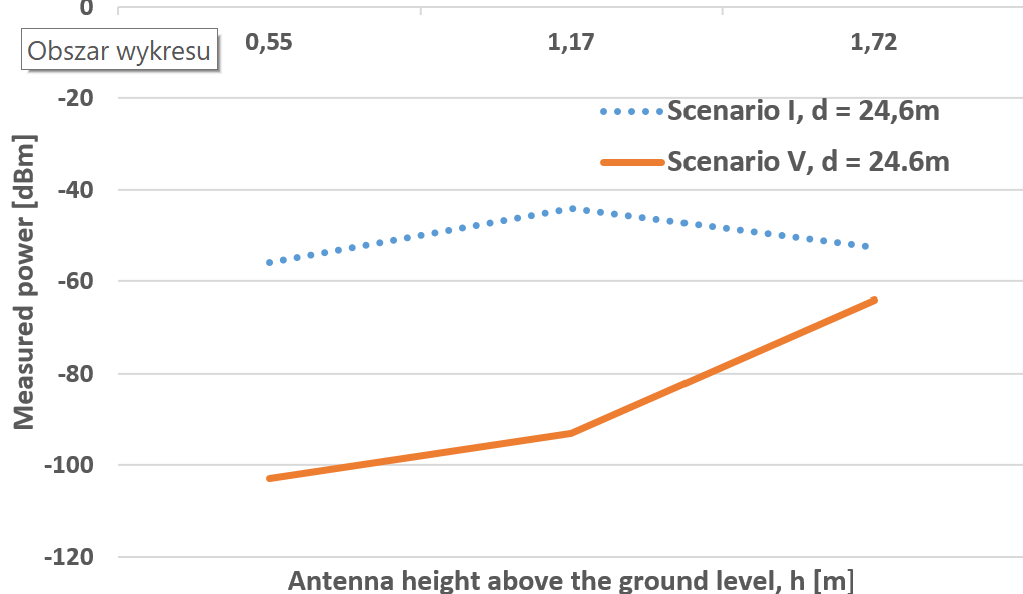}
\caption{Measured power as function of antenna height in Scenarios I and V}
\label{fig_PLfour}
\end{figure}

\section{Conclusion}
In this paper, the results of conducted measurements of channel attenuation at frequency 26.555 GHz have been presented. It has been shown that the presence of even one or two cars can significantly reduce the range of communications at such high frequencies, as the induced signal attenuation due to the presence of blocking cars reaches even 60 dB. On the other side, it was proved that - as expected - there is a high impact of the antenna placement on the observed phenomena. In particular, it seems beneficial to install the antennas at the roof-top level when there is a need to guarantee high communication distance within the platoon. However, in the case of multi-hop transmission (when the messages are forwarded from car to car), the placement at car bumper level could be feasible to minimize the prospective interference. The conducted experiments have been carried out in static scenarios, hence, it is important to further investigate the impact of platoon mobility. Moreover, the observed attenuation will be probably worse when instead of small cars the platoon will be constituted by trucks or lorries.


\section*{Acknowledgment}
The work has been realized within the project no. 2018/29/B/ST7/01241 funded by the National Science Centre in Poland.



\bibliographystyle{IEEEtran}
\bibliography{references.bib}
%




\end{document}